\documentclass[twocolumn,english,aps,prb,twocolum,superscriptaddress,natbib,bibnotes,amsmath,amssymb,floatfix,groupedaddress,footinbib]{revtex4-2}
\pdfoutput=1
\usepackage[colorlinks=true,citecolor=blue,linkcolor=magenta]{hyperref}

\usepackage[markup=nocolor, authormarkupposition=left]{changes} 

\usepackage{soul}
\usepackage[utf8]{inputenc}
\usepackage[english]{babel}
\usepackage{amsmath,amsfonts,amssymb}
\usepackage[T1]{fontenc}
\usepackage{url}

\usepackage{amsmath}
\usepackage{siunitx}
\usepackage[version=4]{mhchem}
\usepackage{amsfonts}
\usepackage{amssymb}
\usepackage{epstopdf}
\usepackage{graphicx}
\graphicspath{{./Figures/}}

\usepackage{changes}

\begin{document}

\title[Article Title]{Single-Shot Integrated Speckle Spectrometer with Ultrahigh Bandwidth-to-Resolution}

\author{Wenzhang Tian$^\ddagger$}
\author{Hao Chen$^\ddagger$}
\author{Mingyuan Zhang}
\author{Zengqi Chen}
\author{Yeyu Tong}
\email{yeyutong@hkust-gz.edu.cn}
\affiliation{$^1$Microelectronic Thrust, \\The Hong Kong University of Science and Technology (Guangzhou), 511453, Guangzhou, Guangdong, PR China\\$^\ddagger$These authors contributed equally to this work.}
\maketitle

\noindent\textbf{\noindent
Miniaturized spectrometers employing chip solutions are essential for a wide range of applications, such as wearable health monitoring, biochemical sensing, and portable optical coherence tomography. However, the development of integrated spectrometers is hampered by the inherent trade-off between bandwidth-to-resolution, footprint, sampling channels, and operation speed. Here, we demonstrate that an ultrahigh bandwidth-to-resolution reconstructive spectrometer can be easily implemented through a single image capture of the speckle pattern diffracted from a passive silicon photonic chip. By leveraging the high pixel count of an image sensor, we can instantly acquire a significant number of distinct spatial sampling channels. Those sampling channels are spatially decorrelated by using our passive optical network on chip including cascaded unbalanced Mach–Zehnder interferometers, random diffraction by an antenna array, and mutual interference in free space before being captured. Hence, each speckle pattern contains wavelength-specific information across its spatial distribution to enhance the effectiveness of the global sampling strategy. Experimentally, we achieve a spectral resolution of 10 pm and an operational bandwidth of 200 nm, with sampling channels up to 2730. Multiple unknown narrowband and broadband spectra can also be precisely obtained.
}

\section*{Introduction} 
In recent decades, the demand for advanced spectroscopic measurement techniques that can be used in situ, in vitro, and in vivo has significantly increased. These developments have been crucial for various applications, such as wearable health monitoring devices\cite{R1,R2}, portable chemical sensors\cite{R3,R3-2}, remote sensing\cite{R3-1,R3-3}, and compact optical imaging systems\cite{R4,R4-1}. This growing trend has driven the rapid advancement of compact optical spectrometers on chip, offering notable benefits in scalability and cost-effectiveness through the use of complementary metal-oxide semiconductor (CMOS) compatible photonics integration platforms. Recent mini or micro prototypes include the development of wavelength demultiplexing spectrometers, which utilize dispersive components, narrowband filters, spatial or temporal heterodyne Fourier transform spectrometers (FTSs) \cite{R5,R6,R7}. Traditional spectrometer construction methods typically rely on spectral-to-spatial or spectral-to-temporal mapping, employing technologies such as arrayed waveguide gratings (AWGs)\cite{R8,R8-1}, echelle gratings (ECGs)\cite{R9,R9-1,R9-2}, dispersive photonic crystals\cite{R10,R11,R12}, tunable ring resonators\cite{R12-1,R23,zhanglong}, and holography\cite{R13}. However, the existing one-to-one mappings between spectral components and spatial or temporal channels are limited, ultimately constraining the bandwidth-to-resolution ratio. Integrated FTSs also encounter a trade-off between performance and power efficiency\cite{R14,R14-1,R14-2,R14-3}, as their resolution is inversely correlated with the optical path length or applied voltage, while their bandwidth is dependent on extensive interferometer arrays or sophisticated wavelength-division-multiplexing (WDM) techniques.
\\\hspace*{1em}To tackle this issue, reconstructive spectrometers based on compressive sensing methodology can be employed. In recent years, miniaturized and programmable photonic circuits have emerged as effective solutions to overcome the bandwidth-to-resolution bottleneck\cite{R17,R26}, demonstrating a record ratio exceeding 20,000\cite{R26}. Such impressive spectroscopic performance was facilitated by a substantial number of sampling channels obtained through reconfigurable and multi-stage tunable filters. As a result, temporally decorrelated optical sampling channels can be aggregated across various on-chip network states. Additionally, the two-dimensional Fourier-transform spectrometer (2D-FTS) presents an innovative approach to overcome the inherent bandwidth-to-resolution limitations by employing a 2D Fourier transform for spectrum reconstruction. This approach significantly improves both scalability and spectral resolution, achieving a resolution of 250 pm over a 200-nm bandwidth, with potential to reach 125 pm through advanced computational methods\cite{R14}. Nevertheless, the resolution of the spectrometer remains subject to trade-offs concerning system complexity, footprint, and sampling speed. On-chip spectrometers continue to present significant challenges, necessitating advancements in bandwidth, resolution, operation speed, and reductions in footprint and cost. 
\\\hspace*{1em}Speckle spectrometers have gained considerable attention due to their ability to achieve high-precision and broadband spectral reconstruction through a single speckle image.\cite{speckle_based_spectrum_analyzers, zhangzunyue}. By leveraging random interference effects in passive photonic structures, these spectrometers map input spectra to complex spatial patterns that serve as unique fingerprints for spectral reconstruction. A wide range of physical mechanisms, such as disordered scattering\cite{R18,Caohui_2016_speckle_optica, Diffractive_speckle_1,Diffractive_speckle_2}, microring lattices\cite{Microring_lattice}, and guided-mode multiplexing\cite{yidan_2020}, have been employed to generate wavelength-dependent speckle responses on-chip. While these systems often exhibit high resolution and compact footprints, a critical limitation lies in the effective number of sampling channels. Although the speckle image may contain hundreds of thousands of pixels, only a statistically independent subset contributes unique spectral information, as spatial correlations introduce redundancy across the speckle field. Consequently, the true spectral dimensionality is not determined solely by the image size, but rather by the number of statistically independent pixels extracted from the speckle pattern. For example, designs that claim several thousand sampling channels may suffer from low channel-to-footprint ratios when correlation between neighboring pixels is high. Thus, maximizing the number of independent sampling channels, rather than merely increasing resolution or bandwidth, has become a key challenge in the design of speckle spectrometers. This requires meticulous engineering of both the spectral encoding process and the spatial structure of the speckle, to ensure minimal redundancy and optimal use of sensor pixels. In this context, a quantitative understanding of spatial decorrelation is essential for evaluating and improving the information throughput of speckle spectrometers.
\\\hspace*{1em}In this study, we propose and experimentally demonstrate an ultrahigh bandwidth-to-resolution reconstructive spectrometer, enabled by a single-shot capture of a speckle pattern from a compact silicon photonic chip. To achieve a substantial number of sampling channels while minimizing power consumption and fabrication costs, a fully passive on-chip random optical network is employed as a wavelength-sensitive encoder. This network comprises cascaded unbalanced Mach-Zehnder interferometers (MZIs) and an antenna array, which emits wavelength-dependent speckle patterns into free space. Unlike reconfigurable photonic chips that rely on sequential sampling through temporal modulation, our system captures a significant number of uncorrelated sampling channels in a single frame using an infrared camera. Each pixel in the captured speckle image represents a potential sampling channel. However, only those pixels that are statistically uncorrelated across the spectral dimension can be considered as independent sampling channels. Through spatial correlation analysis, we estimate the number of statistically independent sampling channels in our system to be approximately 2730 ($\rho_{thr}=0.5$). This substantial number of sampling channels enables high-resolution spectral encoding and reconstruction over a broad bandwidth. In our experimental demonstration, the 2 mm$^2$ passive photonic chip achieves a spectral resolution of 10 pm within a 200 nm bandwidth using only one-shot image acquisition. Moreover, our system successfully reconstructs unknown broadband and bandpass spectra, as well as arbitrary spectra with sharp spikes.
\section*{Speckle spectrometer design}
\begin{figure*}[tb]
  \includegraphics[width=\linewidth]{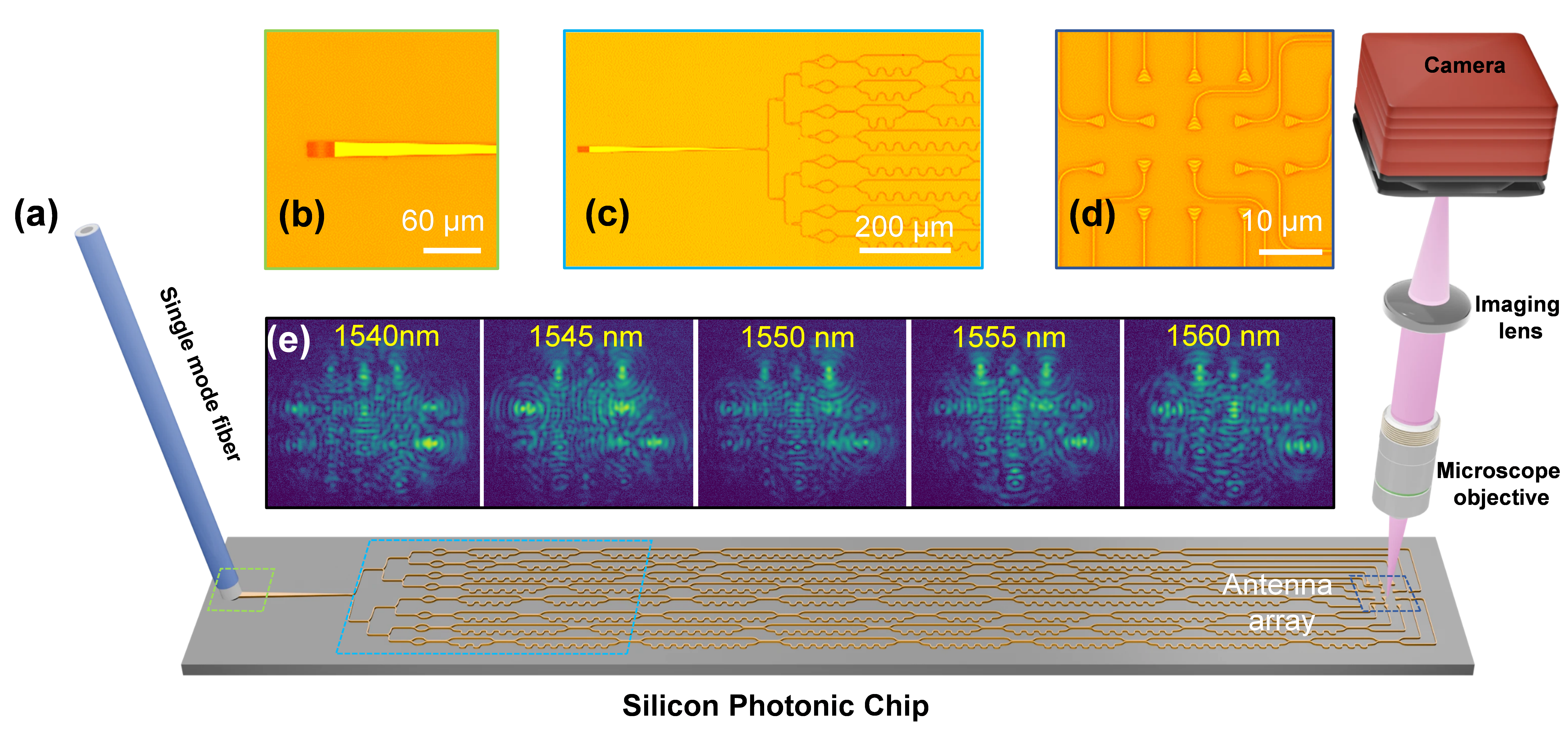}
  \caption{\noindent\textbf{Schematic of the proposed integrated speckle spectrometer.} (a) Layout of the spectrometer. A microscope objective and an imaging lens are utilized to capture speckle images at various spatial position. (b) Speckle images measured at various incident wavelengths. (c) Microscope image of the input broadband grating coupler. (d) Microscope image of unbalanced MZIs-based random optical network. (e) Bright field image of the antenna array.}
  \label{fig:boat1}
\end{figure*}

Figure 1a illustrates our proposed single-shot speckle spectrometer. To fully exploit the advantages of global sampling strategies in speckle spectrometers and achieve ultra-high spectroscopic performance, a large number of broadband sampling channels with optimized spectral responses are crucial. 

In our spectrometer design, we propose a purely passive photonic network comprising multiple stages of unbalanced MZIs and a compact antenna array on a silicon-on-insulator (SOI) platform. The MZI network introduces strong wavelength-dependent phase variations, while the antenna array diffracts the encoded optical signals into free space at wavelength-dependent angles. These two components collectively produce randomized speckle patterns with minimal system complexity and power consumption. Our photonic chip features a 220-nm-thick silicon layer atop a 2-\textmu m-thick buried silicon dioxide layer. The spectral properties of each interferometer are precisely engineered to produce an overlaid transmission spectrum with rapid pseudo-random fluctuations, resulting in a narrow auto-correlation width that improves global sampling efficiency. By randomly tuning the arm length of each interferometer, the resultant transmission spectra become decorrelated. At the end, a compact antenna array diffracts the optical signal out of the chip plane. The wavelength-dependent emission angle of the antenna further enhances the spatial decorrelation of the sampling channels\cite{R37}. We incorporate an imaging system that includes an objective(MY10X-823), an imaging lens(MVL12X3Z) and an infrared camera(ARTCAM-991SWIR) to capture the diffracted speckle pattern from the photonic chip. As a result, the expansion and contraction of the field of view directly influence the number of sampling channels (corresponding to the number of independent pixels in the speckle image), thereby impacting our overall sampling capability. This architecture enables a single image to yield thousands to tens of thousands of effective sampling channels. A limited number of purely passive waveguide components is needed, thus significantly boosting the resolution of the spectrometer, while circumventing the time overhead and on-chip power consumption commonly associated with tunable phase shifters. 

When a specific wavelength of light is coupled into our speckle spectrometer through an input broadband grating \cite{grating}, as depicted in Figure 1b and Figure S1a-d (see more details about the design parameters in Supplementary Note 1), it enters a cascaded network of unbalanced MZIs. Figure S1a illustrates the schematic of the input grating, while S1b shows the corresponding scanning electron microscopy (SEM) image with detailed geometric parameters. The near-field emission profile and simulated transmission performance of the grating are depicted in Figure S1c and S1d, respectively, demonstrating efficient broadband coupling across the 1400–1600 nm range. Once the light is coupled into the chip, it propagates through a cascaded network of random unbalanced MZIs, introducing wavelength-dependent phase delays along different optical paths, as depicted in Figure 1c. The accumulated dispersion resulting from these phase delays significantly influences both the intensity and phase distributions, leading to wavelength-specific interference patterns. Subsequently, the light radiates through the antenna array, as shown in Figure 1d and Figure S2a-S2f (see more details about the design parameters in Supplementary Note 1), where further wavelength-dependent free-space interference enhances the speckle complexity. This produces distinct speckle patterns that shift dynamically with wavelength changes. Finally, the resulting wavelength-dependent speckle patterns are captured as unique optical fingerprints by a microscope objective (MY10X-823), an imaging lens (MVL12X3Z), and an infrared camera (ARTCAM-991SWIR), as shown in Figure 1e and Figure S3, facilitating accurate spectral reconstruction (see more details about the measurement method in Supplementary Note 2).
\\\hspace*{1em}As illustrated in Figure 2a, the proposed speckle spectrometer relies on a carefully engineered random optical network architecture to achieve ultrahigh resolution. Our passive on-chip photonic network consists of four key functional regions: the split region, the interference region, the routing region, and the diffraction region. All of these regions jointly manipulate the optical path and phase of light. The split region, formed by 8 cascaded power splitters, distributes the input light into multiple optical paths, initiating the wavelength-dependent interference. The split light is then recombined in the interference region, where randomized optical path difference and phase diversity are introduced. These factors are critical for generating highly resolved speckle patterns. The arm length in the unbalanced MZIs are designed with randomized optical path differences, generated using a random number seed within the range of 100 to 500 \textmu m, with each stage consisting of 8 cascaded interferometers. The routing region further enhances the optical path difference by redirecting light to the antenna array. The diffraction region comprises 16 evenly spaced antennas, each connected to the routing region via 500 nm-wide waveguides incorporating bends with a radius of 10\textmu m. The orientation of each antenna is varied to further accelerate the decorrelation of speckle patterns with respect to wavelength. In this region, light is diffracted into free space at varying angles, introducing additional wavelength-dependent variations. 
\\\hspace*{1em}The exhibited dispersion, interference, and diffraction by the photonic network ensure that the speckle patterns are highly decorrelated and sensitive to wavelength variations. Additionally, the antenna array facilitates free-space radiation and interference, further enhancing the generation of high-resolution, wavelength-dependent speckle patterns that are captured by the image sensor. These patterns, serving as highly distinctive optical signatures, enable precise and efficient spectral reconstruction.

\begin{figure*}[t]
  \includegraphics[width=\linewidth]{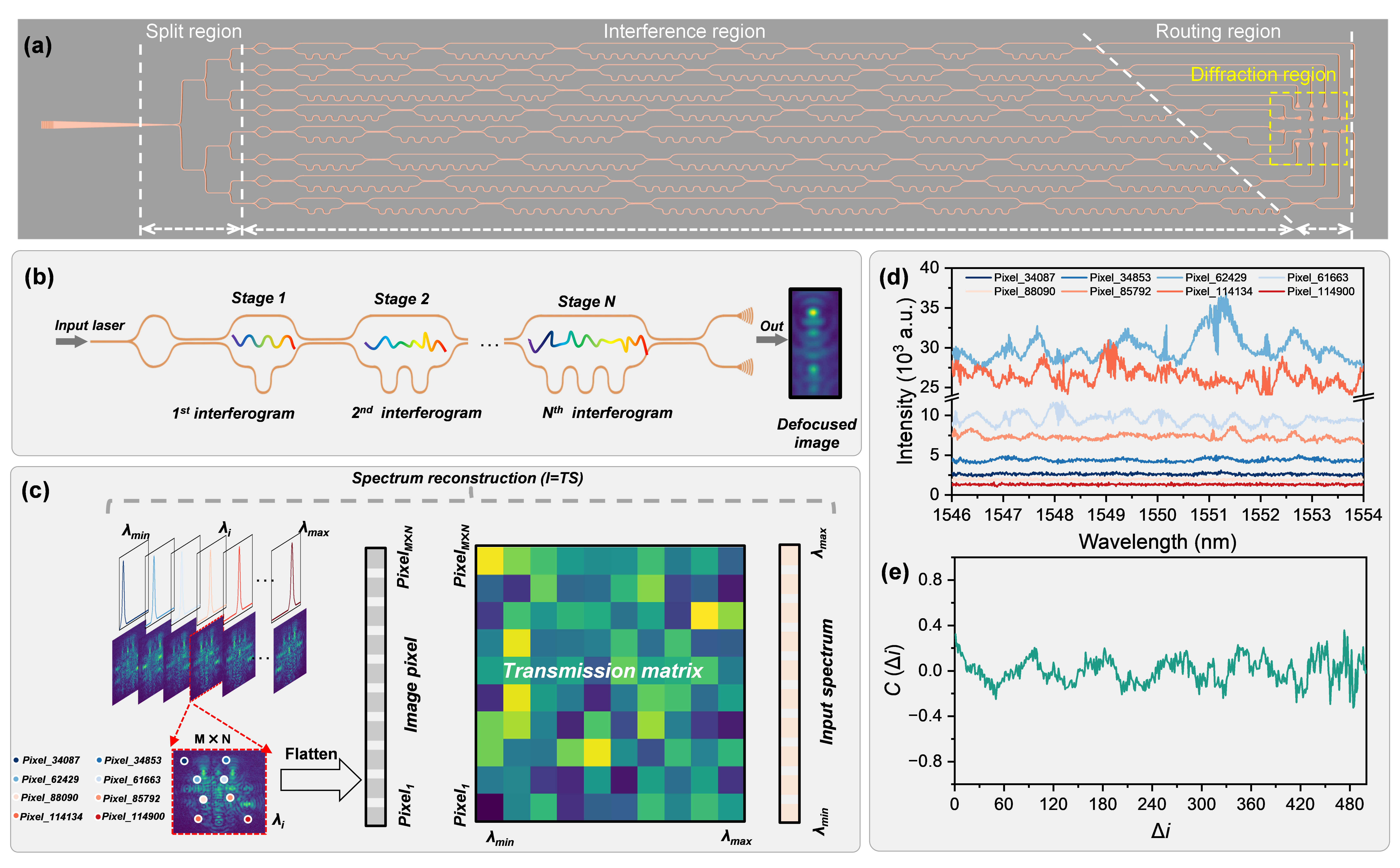}
  \caption{\noindent\textbf{Design and principle of the speckle spectrometer.} (a) Schematic layout of the proposed speckle spectrometer, consisting of an input broadband grating coupler, a split region, an interference region composed of multiple stages of unbalanced MZIs, a routing region, and a diffraction region with an antenna array. (b) The cascaded transmission spectrum formed by the superposition of multiple unbalanced MZI stages. (c) Spectral reconstruction mechanism. The 2D speckle image captured by the infrared camera is flattened into a 1D intensity vector \(I \in \mathbb{R}^{MN}\), where \( \mathbb{R} \) denotes the space of real-valued entries, \(M\) and \(N\) are the image dimensions. The unknown input spectrum is denoted as \(S\), and the calibrated transmission matrix as \(T\). The inverse problem \(I = TS\) is solved to recover \(S\). The inset shows spectral decorrelation across channels. (d) Spatial response \( \mathrm{Pixel}\_{i} \) versus wavelength \(\lambda\). (e) Spectral cross-correlation coefficient \(C(\Delta i)\) computed between speckle patterns at different wavelengths, where \(\Delta i\) denotes the index offset between two wavelengths. Lower values of \(C(\Delta i)\) indicate stronger spectral decorrelation across wavelength channels.}
  \label{fig:boat1}
\end{figure*}
\section*{Operation Principle} 
Figure 2b schematically illustrates the implementation principle of our system via a series of unbalanced MZIs. In our system, the transmission spectrum of an arbitrary MZI stage \(T_i\), after pass through the \(N_{th}\) interferogram, can be described as:
\begin{equation}
T_{i} = \prod I_{i} \cos(\beta \Delta L + \delta_{i})
\end{equation}
\\where the amplitude of the periodic oscillation term, $\cos(\beta \Delta L + \delta)$, is influenced by the power splitting ratio of the power splitter, as well as the input power \(\textit{I}_{i}\). Meanwhile, the phase of the signal is determined by the optical path difference $(\Delta L)$  between the two arms of the interferometer, along with an additional relative phase shift $(\delta)$  introduced by external factors such as reflection or refraction. By tailoring the spectral response of each MZI stage, the overall transmission spectrum can be fully manipulated. Specifically, we strategically adjust $\Delta L$ in the arms of various MZI stages to disrupt the inherent periodicity of individual interferogram. As a result, the combined spectral response exhibits pseudo-random variations across the wavelength range. 

Arbitrary unknown input spectra, denoted as $\textit{S}({\lambda})$, propagate through the cascaded optical network with a spectral response \(\textit{T}({\lambda})\)\ , and subsequently emitted into free space via the antenna array. Due to mutual interference among the optical fields emitted from multiple antennas, complex speckle patterns are generated. These wavelength-dependent speckle patterns are then captured by an infrared camera and serve as spatially encoded signatures of the input spectrum. Each speckle pattern represents a unique spatial intensity distribution resulting from the collective interference effects of all antennas at a specific input wavelength. Simultaneously, the 2D speckle patterns captured at different wavelengths are flattened into 1D vectors, which are then combined to construct a response output matrix $I$. This systematic approach efficiently encodes the spectral information, providing a foundation for accurate and effective spectral reconstruction. It can be described as follows:
\begin{equation}
I=\int_{\lambda_{min}}^{\lambda_{max}} T(\lambda)S(\lambda) \, d\lambda
\end{equation}

\noindent where \(\lambda_{\text{min}}\) and \(\lambda_{\text{max}}\) represent the lower and upper limits of the operating wavelength, respectively. Considering an image with \( M \times N \) pixels, the captured speckle pattern can be flattened into a vector and modeled as:
\begin{equation}
I_{(M \times N) \times 1} = T_{(M \times N) \times L} S_{L \times 1}
\end{equation}
Here, each pixel in the speckle image (consisting of $512\times 512=262,144$ pixels) is mathematically considered as a potential sampling channel. However, due to spatial correlations within the speckle field, many neighboring pixels convey redundant information. As a result, only a subset of these pixels are statistically independent and effectively contribute non-redundant information, thus determining the actual number of sampling channels utilized for accurate spectral reconstruction. Additionally, each element of \textit{I} represents the intensity captured by a single pixel in the 2D speckle image at a given input wavelength. The transmission matrix \textit{T} is obtained through pre-calibration (see detailed algorithm 1 in Supplementary Note 3), and \textit{S} is the input unknown spectrum. The degree of decorrelation among spectral channels in the transmission matrix critically influences the quality of spectral reconstruction, with higher decorrelation promoting greater spectral distinguishability. Figure 2c illustrates the calibrated transmission matrix \textit{T}, in which the color map encodes the transmittance intensity. Each row exhibits a distinct, non-periodic response pattern, highlighting the intrinsic complexity of the transmission system. To quantitatively evaluate the spectral encoding performance of the transmission matrix \textit{T}, we perform a correlation analysis on \textit{T}. This analysis characterizes the degree of decorrelation among spectral channels, which reflects the capability of the system to perform accurate spectral reconstruction. The correlation function of the transmission matrix \textit{T} is defined as follows:
\begin{equation}
C(\Delta\lambda,N) = \frac{\langle I(\lambda,N) \rangle \langle I(\lambda+\Delta\lambda,N) \rangle}{\langle I(\lambda,N) I(\lambda+\Delta\lambda,N) \rangle} - 1
\end{equation}
\\where \(\mathit{I}(\lambda, N)\) denotes the flattened speckle intensity vector at wavelength $\lambda$, and $\Delta\lambda$ is the spectral offset. The averaging operator $\langle\cdot\rangle$ ensures statistical robustness by averaging over the entire set of sampled wavelengths. Figure 2d presents the spectral responses of eight randomly selected pixels from a single speckle image, each representing a potential sampling channel. The responses exhibit pronounced spectral fluctuations, highlighting the high degree of randomness in the wavelength domain. This randomness is critical for ensuring effective spectral encoding and reconstruction. Furthermore, as shown in Figure 2e, the correlation coefficients between entire speckle images at different wavelengths remain consistently below 0.5, highlighting strong inter-image decorrelation across the spectral domain. This low cross-correlation demonstrates the effectiveness of the transmission matrix in maintaining channel independence, which is crucial for achieving accurate spectral reconstruction. Finally, to evaluate the performance of spectral reconstruction, we employ the relative error ($\varepsilon$) metric, which quantifies the deviation between the reconstructed spectrum $\hat{S}$ and the reference spectrum $\bar{S}_0$:
\begin{equation}
\varepsilon = \frac{|| \hat{S}- \bar{S}_0||_{2}}{||\bar{S}_0||_2}
\end{equation}
where $||\cdot||_2$ denote the $l_{2}$ norm, $\hat{S}$ denotes the reconstructed spectrum, and $\bar{S_0}$ is the reference spectrum.

\begin{figure*}[t]
  \centering
  \includegraphics[width=1\linewidth]{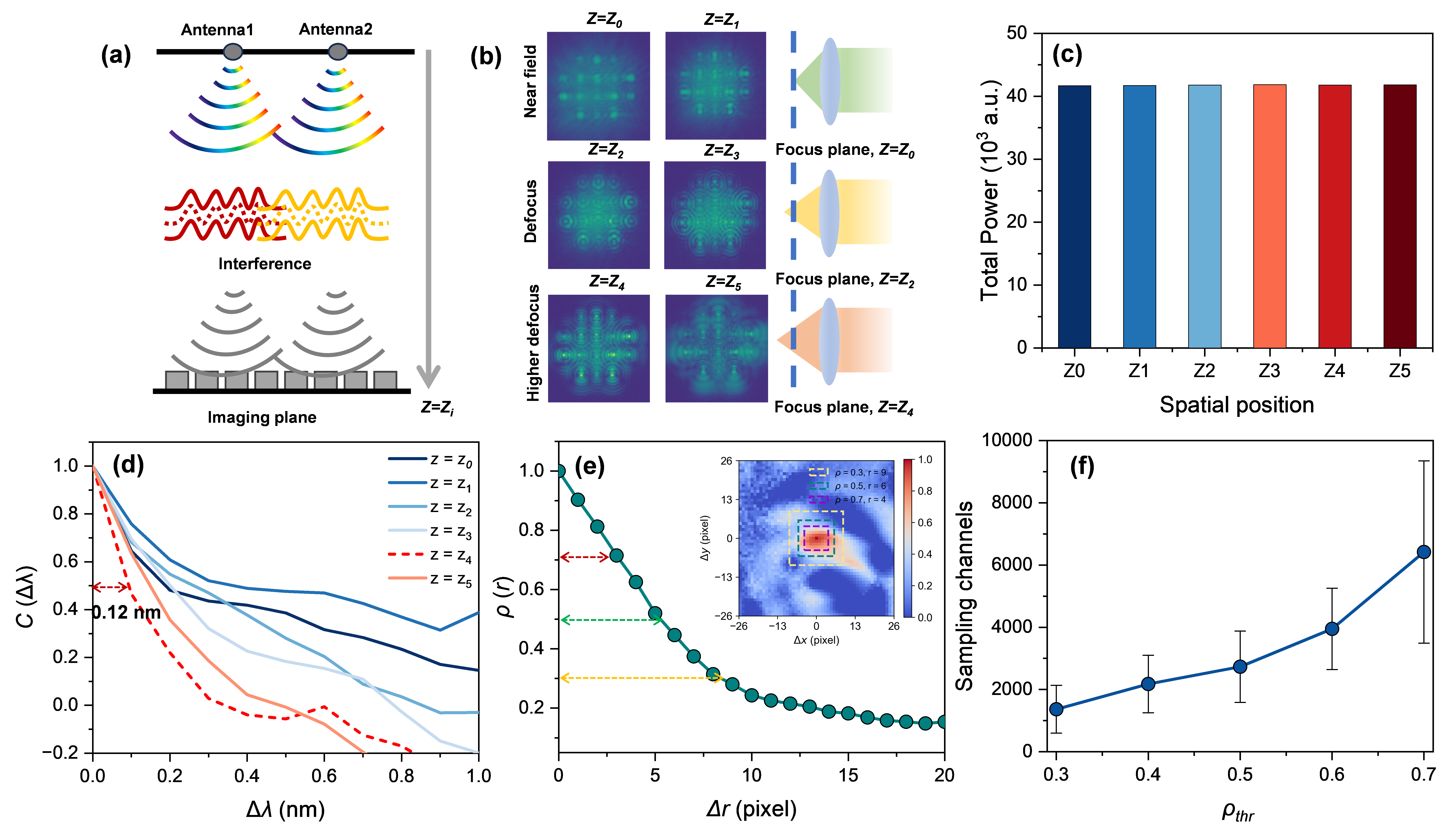}
  \caption{\noindent\textbf{Characterization of the speckle spectrometer.} (a) Schematic of wave diffraction from two antennas. Light is diffracted into free space at a wavelength-dependent angle, generating a complex speckle pattern on an imaging plane at a depth of $z_{i}$. (b) Speckle images were captured with the imaging plane positioned at three distinct locations. (c) Total optical power (i.e., the sum of pixel intensity) of different spatial positions ($z_{i}$). (d) The spectral correlation coefficients of light intensities averaged over the entire wavelength range for various $z_{i}$. The red dashed line indicates a half-width at half-maximum (FWHM) of 0.12 nm, meaning that a wavelength shift of 0.12 nm reduces the spectral correlation to half. This narrower FWHM suggests that $z_{4}$ offers the optimal spatial position to reconstruct spectrum. (e) Estimation of the minimum uncorrelated radius ($r_{min}$). The inset shows the Pearson correlation coefficient between the central pixel and its adjacent pixel. (f) Estimated number of sampling channels of speckle image from 1540 nm to 1560 nm. Specifically, at $\rho_{thr}=0.5$, we obtain around 2730 sampling channels.}
  \label{fig:boat1}
\end{figure*} 
\begin{figure*}[!t]
  \includegraphics[width=\linewidth]{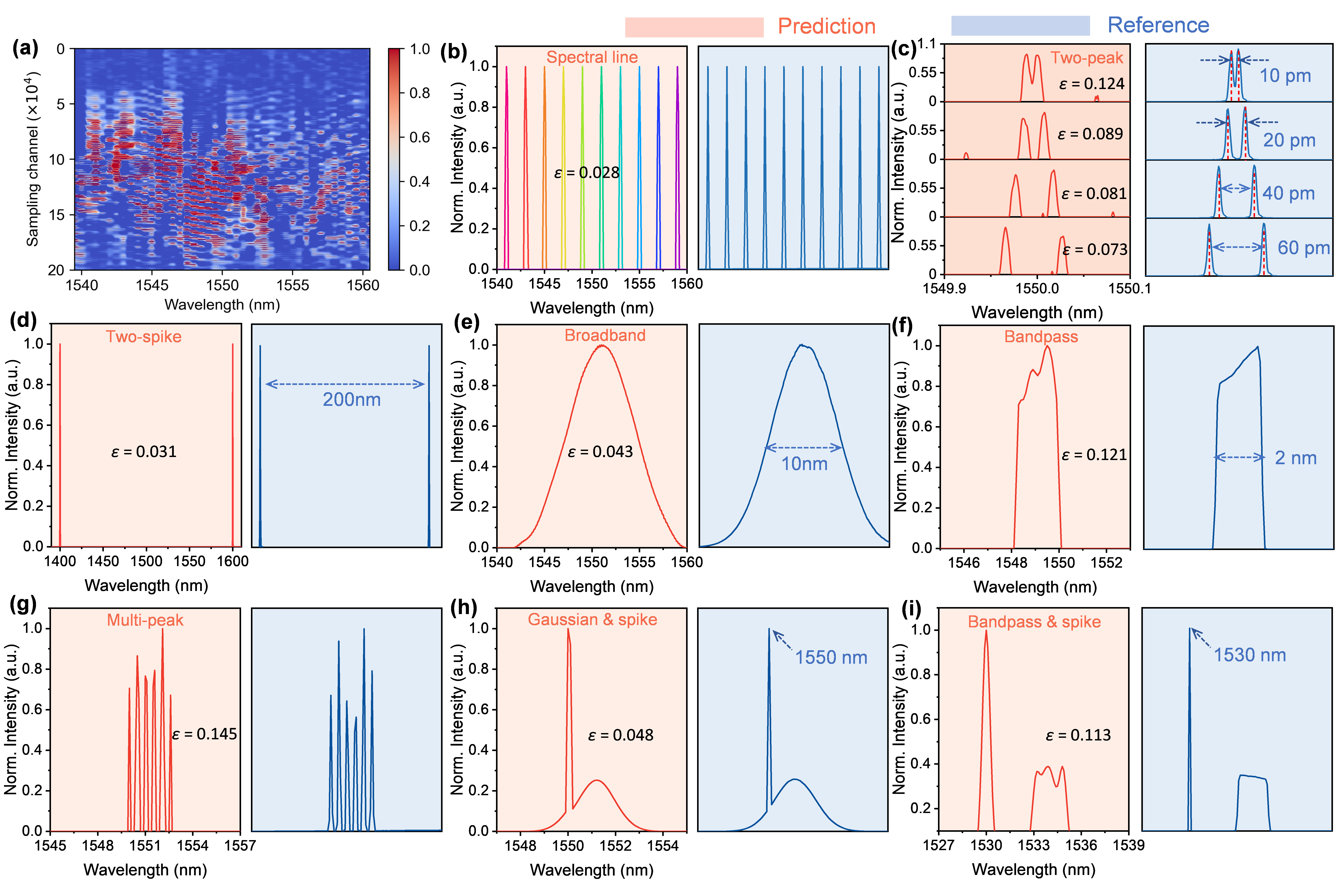}
  \caption{\noindent\textbf{Spectral calibration and testing of the speckle spectrometer.} (a) Calibrated transmission matrix \( T \), encoding the mapping between input wavelengths and speckle intensity distribution \(I\) across the image sensor pixels. $T$ is obtained via pre-calibration using a tunable laser source (TLS) by capturing speckle images at discrete wavelengths. (b) Reconstructed spectrum \(\hat{S}\) for a series of narrowband probe signals over a 20 nm range. Blue lines denote the true center wavelengths of the probe, and the rainbow line represents the reconstructed spectrum. The reconstructed spectral linewidth exhibits a full width at half maximum (FWHM) of less than 0.1 nm. (c) Reconstructed spectra of two closely spaced spectral lines with varying separation, used to quantify the resolution limit of the spectrometer. (d) Reconstructed spectrum demonstrating an operational bandwidth of 200 nm. (e) Reconstructed spectrum for a continuous, broadband probe spectrum with FWHM = 10 nm. (f) Reconstructed spectrum for an amplified spontaneous emission (ASE) source pass a 2 nm bandpass filter. (g) Reconstructed spectrum consisting of six closely spaced peaks, each with a FWHM of 0.2 nm. (h) Reconstructed spectrum for a Gaussian spectrum (FWHM = 3 nm) with an added spike. (i) Reconstructed spectrum for a spike combined with the ASE spectrum, after being filtered by a 2 nm bandpass filter.}
  \label{fig:boat1}
\end{figure*}
\section*{Results}
As illustrated in Figure 3a, light radiated from two antennas undergoes interference during propagation through free space, producing complex speckle patterns. These patterns are collected via an objective lens and an imaging lens, and are recorded by an infrared camera. The resulting speckle patterns observed at various spatial positions \(z_{i}\), as illustrated in Figure 3b, reflect the interference between radiated waves at different spatial positions. In the near field region, specifically at positions \(z_{0}\) and \(z_{1}\), the interference between the mutually coupled antennas is relatively weak. This limits the optical path difference between radiated waves of varying wavelengths, resulting in insufficient decorrelation of the speckle patterns. Consequently, the speckle patterns at \(z_{0}\) and \(z_{1}\) exhibit minimal variation with wavelength as evidenced by the relatively similar structures observed in Figure 3b. To overcome this limitation and increase the optical path difference between different wavelength components, the imaging plane was intentionally defocused. At these defocused positions, where the optical path lengths between wavelengths differ significantly, the pronounced changes in the speckle patterns lead to greater spectral decorrelation. The speckle patterns become increasingly sensitive to variations in wavelength, as reflected in the distinct and diverse speckle patterns observed in the defocused images. This highlights the enhanced optical path difference achieved through defocusing. Figure 3c presents the total power emitted by the antenna array, as recorded by the infrared camera at various spatial positions \(z_{i}\). The relatively uniform total power across all positions confirms consistent radiation from the antennas, ensuring that the speckle patterns are unbiased by any variations in power distribution. To quantify the degree of spectral decorrelation at different spatial positions, Figure 3d presents the calculated correlation function \(\mathit{C}(\Delta\lambda)\), which characterizes how quickly the speckle patterns decorrelate with wavelength shifts $\Delta\lambda$, for various \(z_{i}\) positions. Notably, the red dashed line in Figure 3d indicates that at position \(z_{4}\), the spectral correlation length drops to 0.12 nm, representing the strongest spectral decorrelation among the six spatial positions analyzed. This result demonstrates that \(z_{4}\) provides the greatest optical path difference, which enhances spectral decorrelation. Consequently, this position enables high-resolution and accurate spectral reconstruction. Furthermore, the spatial position \(z_{i}\) of the infrared camera plays a critical role in ensuring accurate spectral measurements. Any deviation in the spatial position of the camera alters the spatial sampling of the speckle patterns, which can invalidate the pre-calibrated transmission matrix, necessitating recalibration. To ensure optimal performance, the infrared camera was fixed at the \(z_{4}\) position, where the enhanced optical path difference supports robust and precise spectral reconstruction. 
We additionally estimate the number of statistically independent sampling channels by using spatial correlation analysis (see details in Supplementary Note 4). A composite matrix $I \in \mathbb{R}^{MN \times L}$ is first formed by concatenating \(L\) speckle images of dimension \(M \times N\) ($M=512$,$N=512$), where each column corresponds to a different spectral channel. We then calculate the Pearson correlation coefficient \(C(x,y)\) between the spectral response at a selected reference pixel and those of its surrounding pixels. To characterize the spatial scale of statistical independence, we derive the radial correlation function $\rho (r)$ by averaging \(|C(x,y)|\) over concentric annuli centered at the reference pixel. The minimum uncorrelated radius ($r_{\min}$) is defined as the smallest radius at which $\rho (r)$ drops below a prescribed threshold $\rho_{\mathrm{thr}}$, as illustrated in Figure 3e. The inset of Figure 3e shows the spatial correlation map, with overlaid colored boxes representing the estimated independent sampling units for different thresholds $\rho_{\mathrm{thr}}$, where each box has a side length of $2r_{\min}$. Accordingly, the total number of statistically independent sampling channels can be approximated by dividing the total image area $A_{total}=M\times N$ by the area of a single independent sampling units, yielding ${\text{Sampling channels}}=A_{total}/(2r_{min})^2$. This result represents the number of spatially independent sampling channels in the composite speckle dataset across all spectral channels. Figure~3f illustrates the estimated number of statistically independent sampling channels as a function of the correlation threshold \(\rho_{\mathrm{thr}}\). As \(\rho_{\mathrm{thr}}\) increases from 0.3 to 0.7, the number of independent sampling channels rises from approximately 1,360 to 6,420. Notably, at \(\rho_{\mathrm{thr}} = 0.5\), the number of independent sampling channels is estimated to be around 2,730. These values are computed from a composite speckle matrix constructed by stacking all measurements acquired over the 1540-1560 nm spectral range, capturing the combined spatial-spectral information. The high spatial decorrelation of the speckle field, enabled by the on-chip random optical network, plays a key role in maintaining a large number of statistically independent measurements. Without sufficient spatial randomness in the optical field, most image pixels would exhibit high correlation, leading to significant redundancy and a substantial reduction in effective sampling channels.
\\\hspace*{1em}Figure 4a illustrates a representative transmission matrix, which serves as the foundation for spectral reconstruction. After pre-calibration, an arbitrary probe spectrum \(S\) can be reconstructed by measuring the intensity \(I\) at the infrared camera and applying the pseudo-inverse of the calibrated transmission matrix as \(\hat{S} = T^{\dagger} I\). In practice, the matrix inversion process is highly sensitive to experimental noise (as shown in Supplementary Figures S4a and S4c), which can significantly degrade the accuracy of spectral reconstruction\cite{speckle_based_spectrum_analyzers}. To mitigate this challenge and improve the reliability of the reconstruction process, we implemented a hybrid approach. Firstly, the pseudo-inverse solution is refined through a nonlinear optimization procedure that minimizes the least-squares error term \(\min\left\|I - T\hat{S} \right\|^2\), thereby suppressing noise and enhancing reconstruction fidelity. This gradient-based update ensures more robust spectral estimation, particularly under noisy measurements or ill-conditioned matrices. Following optimization, we further process the result in the discrete cosine transform (DCT) domain to enhance sparsity. Although the original spectrum is not sparse in the spatial domain, applying a DCT converts it into a sparse representation. This sparsity enables the application of compressed sensing techniques~\cite{R20,R20-1,R20-2}, which allow efficient reconstruction of spectral information using significantly fewer measurements—a key advantage in high-resolution spectral sensing. Finally, the optimized result is treated as a sparse spectrum \(\hat{S}_{\mathrm{DCT}}\), and the inverse discrete cosine transform (IDCT) is applied to recover the original spectrum: \(\hat{S} = \mathrm{IDCT}(\hat{S}_{\mathrm{DCT}})\). Mathematically, this corresponds to redefining the error term in the DCT domain as \( \left\|I - T_{\mathrm{DCT}} \cdot \hat{S}_{\mathrm{DCT}} \right\|^2\), where \( T_{\mathrm{DCT}} = T \cdot D^{-1} \) and \(\hat{S}_{\mathrm{DCT}} = D \cdot \hat{S}\), with \(D\) being the real orthonormal DCT basis matrix (see algorithm 2 in Supplementary Note 3, Figure S4b, and S4d for details). This transformation enhances the sparsity of the spectrum, making compressed sensing techniques applicable and effective. As a result, it enables accurate reconstruction of spectral lines from fewer measurements, while improving robustness to experimental noise.
In our current implementation, both transmission matrix calibration and spectral reconstruction are executed efficiently on a standard desktop computer (CPU: 13th Gen Intel Core i5-13600KF). Specifically, image acquisition for each spectrum requires only 0.5 ms, enabling rapid data collection. The calibration of the transmission matrix, a one-time initialization step to establish the spectral response mapping of the system, takes approximately 4 s. Once the transmission matrix is pre-calibrated, each unknown spectrum can be reconstructed in approximately 0.55 s.
\\\hspace*{1em}Based on the combined matrix inversion and nonlinear optimization approach described above, we systematically evaluated the performance of the speckle spectrometer under various test conditions. Firstly, a series of tunable laser signals at different wavelengths were launched as narrow-band inputs. These signals were selected for their well-defined spectral characteristics, providing a reliable baseline for assessing the reconstruction accuracy of the spectrometer. Figure 4b shows the reconstructed laser signals in the DCT domain, demonstrating excellent agreement with the reference spectra. The full width at half maximum (FWHM) of the reconstructed peaks at various wavelengths is consistently around 10 pm, indicating that the system is capable of reconstructing fine spectral features with high precision. Next, to further evaluate the resolution limit of the spectrometer, we simultaneously launched two closely spaced laser signals as dual-peak inputs. As the wavelength separation was gradually reduced from 60 pm to 10 pm, as illustrated in Figure 4c, we observed that all peaks remained clearly distinguishable even at a minimum separation of 10 pm, demonstrating the system capability of achieving ultra-high resolution. This level of resolution is crucial for applications requiring the differentiation of closely spaced spectral lines, such as high-resolution spectroscopy or telecommunications. Finally, the broadband capability of the spectrometer was evaluated by reconstructing widely separated spectral components located at 1400 nm and 1600 nm, as shown in Figure 4d. The results confirm a total operational bandwidth of 200 nm, spanning the S, C, and parts of the E and L bands, while maintaining high spectral resolution across the entire range.
\\\hspace*{1em}In addition to narrowband features, we further evaluated the capability of the spectrometer to reconstruct continuous broadband spectra, which were generated using the amplified spontaneous emission (ASE) signal from an erbium-doped fiber amplifier (EDFA) and shaped by either a Gaussian or a bandpass filter, as shown in Figures 4e and 4f. Figure 4e presents the reconstructed spectrum for a Gaussian-shaped broadband signal with a FWHM of 10 nm, which was well recovered with high fidelity. Figure 4f shows the reconstructed spectrum after applying a 2 nm bandpass filter to the ASE signal. However, minor deviations are observed relative to the reference, as shown in Figure 4f. These discrepancies are mainly attributed to the limitations of DCT-based reconstruction, where closely spaced frequency components may only be approximately captured, leading to slight oscillations in the reconstructed spectrum\cite{DCT}. Additionally, we assessed how effectively the spectrometer reconstructs complex multi-peak spectra. As shown in Figure 4g, the spectrometer successfully reconstructs a spectrum composed of six spike spectral lines, evenly spaced between 1545 nm and 1557 nm, each with a FWHM of 0.1 nm. The reconstructed result closely matches the reference spectrum, with all peaks accurately located and clearly resolved. Although minor fluctuations in peak intensity are observed, mainly due to the sensitivity of the DCT-based reconstruction to closely spaced components, the spectral positions remain consistent with the ground truth\cite{DCT}, indicating that the resolution is not compromised. Finally, the performance of the spectrometer in reconstructing hybrid spectral features combining both broadband and narrowband components was assessed. As shown in Figure 4h, the spectrometer accurately reconstructs a spectrum consisting of a smooth Gaussian background (FWHM = 3 nm) superimposed with a sharp spectral spike, highlighting its capability to preserve fine features without sacrificing broader shape integrity. Moreover, Figure 4i presents the reconstruction of a 2-nm-wide bandpass-filtered spectrum superimposed with a narrow, high-intensity spike centered at 1530 nm. To quantitatively assess the reconstruction performance across various spectral inputs, the relative errors \( \varepsilon \) are shown in Figure~4. For simple spectral profiles such as spikes or smooth broadband shapes (Figure~4b and 4e), the relative errors remain below 0.05, demonstrating high reconstruction fidelity. In Figure~4c, which evaluates the resolution limit using two closely spaced peaks, the relative errors vary from 0.073 to 0.124 as the peak spacing decreases, highlighting the increasing challenge of resolving features at smaller separations. In contrast, more complex inputs, including multi-peak spectra (Figure~4g) or hybrid signals combining broad and narrow features (Figures~4h and 4i), yield moderately higher errors up to 0.15. These variations reflect the inherent difficulty in recovering dense or non-sparse spectral content. Nevertheless, across all tested scenarios, the relative error remains below 0.15, demonstrating the robustness and versatility of the proposed reconstruction approach under diverse input conditions. Further analysis of the reconstruction sensitivity under added noise conditions is provided in Supplementary Figure S4.

\begin{table*}[t]
\centering
 \caption{Table 1 Performance comparison with state-of-the-art integrated spectrometers.}
 \resizebox{\textwidth}{!}{
  \begin{tabular}{ccccccccc}
    \hline
    \textbf{Spectrometer} & \textbf{Footprint} & \textbf{Resolution} & \textbf{Bandwidth} & \textbf{Channel}  & \textbf{Operation} & \textbf{Sampling}& \textbf{Fabrication}\\
    \textbf{type} & \textbf{($\mathbf{mm^2}$)} & \textbf{(pm)} & \textbf{(nm)} & \textbf{numbers} & \textbf{complexity} & \textbf{time} & \textbf{platform}\\
    \hline
    Disordered medium\cite{R18} &$1.25\times10^{-3}$ & 750 & 25 & 8 & N.A. & One-shot& Silicon\\
    
    Photonic lantern\cite{R18-1} & 3 & 300 & 5.6 & 64 & High & N.M.& Silicon\\
    
    Meta-lenses\cite{R18-0}&0.01& 140 & 35 & 32 & N.A. & One-shot& Silicon\\
    
    Coherent network\cite{R21}  & 0.1144 & 20 & 12 & 64 & High & One-shot& Silicon Nitride\\
    
    Inverse design 1\cite{R22} & 0.0264 & 100 & 120 & 25 & Low & One-shot& Silicon\\
    
    Inverse design 2\cite{Inverse_design_2} & 5 & 120 & 120 & 64 & Low & N.M. & Silicon Nitride \\

    Tailored Disorder~\cite{Tailored_Disorder} & $3.84\times10^{-4}$ & 250 & 30 & 23 & High & N.M. & Silicon \\
 
    Microdisk resonator\cite{R23} & 0.04 & 200 & 20 & 1 & High & N.M.& Silicon\\
    
    Single resonator\cite{R24} & $7\times10^{-4}$ & 80 & 100 & 1250 & High & $< 0.3s$&Silicon\\
    
    Straified waveguide filter\cite{R25}  & $9.1\times10^{-3}$ & 450 & 180 & 32 & N.M. & N.M.& Silicon\\
    
    2D Fourier transform\cite{R14} & 8.75 & 250 & 200 & 801 & High & $< 0.025s$& Silicon\\ 
    
    Random speckle\cite{R17} & N.M. & 50 & 35 & 3501 & High &$< 5s$ & Silicon\\
    
    Chaotic Reflection~\cite{Chaotic_reflection} & $2.5\times10^{-4}$ & 20 & 90 & 32 & Low & N.M. & Silicon \\

    Diffractive speckle 1~\cite{Diffractive_speckle_1} & 0.045 & 47 & 40 & 851 & Low & One-shot & Silicon \\

    Diffractive speckle 2~\cite{Diffractive_speckle_2} & 0.1425 & 70 & 100 & 1400 & Low & One-shot & Silicon \\

    Microring lattice\cite{Microring_lattice} & 1 & 15 & 40 & 2666 & Low & $0.1\mu\mathrm{s}$ & Silicon \\

    Crossing microring \cite{R16} & 15.2 & 30 & 115 & 256 & High &$< 0.3s$ & Silicon Nitride\\ 
    
    Cascaded microring~\cite{Cascaded_microring} & 0.012 & 8 & 520 & 512 & High & N.M. & Silicon Nitride \\

    MZIs network\cite{R26} & 7.03 & 10 & 200 & 729 & High & $< 0.8s$ & Silicon Nitride\\ 
    
    \textbf{This work} & \textbf{2} & \textbf{10} & \textbf{200} & \textbf{2730 ($\rho_{thr}=0.5$)} & \textbf{Low} & \textbf{One-shot}& \textbf{Silicon}\\
    \hline
    *N.A.: Not applicable; N.M.: Not mentioned.
  \end{tabular}
  }
\end{table*}

\section*{Discussion}
We compared the proposed spectrometer with several prior works based on key performance indicators, as summarized in Table~1. Previous studies, such as those based on disordered media, meta-lenses, and photonic lanterns, are confined to resolutions of 750~pm, 300~pm, and 140~pm, respectively~\cite{R18,R18-1,R18-0}. Although these designs excel in achieving exceptionally compact footprints, a reduced spectral resolution was realized. In contrast, coherent network and random speckle spectrometers offer improved resolutions of 20~pm and 50~pm, respectively~\cite{R21,R17}. However, their bandwidths are limited to 12~nm and 35~nm, with sampling channel counts restricted to 64 and 3501. A common strategy to increase the number of sampling channels in random speckle spectrometers is to employ tunable thermo-optic phase shifters, yet this approach introduces additional operational complexity and slower acquisition times. Similarly, microdisk resonator and single resonator spectrometers offer compact footprints but are limited by resolutions of 200~pm and 80~pm, with sampling channel capacities of only 1 and 1250, respectively~\cite{R23,R24}. Recent spectrometer designs, including inverse-designed~\cite{Inverse_design_2}, tailored-disorder~\cite{Tailored_Disorder}, chaotic reflection~\cite{Chaotic_reflection}, and diffractive speckle-based architectures~\cite{Diffractive_speckle_1,Diffractive_speckle_2}, aim to minimize device footprint while supporting passive operation through engineered scattering or structural disorder. Notably, the chaotic reflection spectrometer achieves the smallest reported footprint of $2.5\times10^{-4}$~mm$^2$, while offering a fine spectral resolution of 20~pm. Nevertheless, these systems typically exhibit limited sampling channels, often fewer than 1500, thereby significantly constraining their capacity for high dimensional spectral reconstruction. The tailored disorder spectrometer provides only 23 sampling channels. Although the chaotic reflection design achieves 20~pm resolution and 90~nm bandwidth, it offers merely 32 usable sampling channels. Cascaded microring and microring lattice spectrometers~\cite{Cascaded_microring,Microring_lattice} provide improved scalability, supporting several hundred to a few thousand sampling channels. However, their resolution (e.g., 15~pm for the microring lattice) or limited real-time performance may hinder their applicability in high-throughput and broadband spectral reconstruction. The reconstruction-based spectrometer employing a MZI network achieves a competitive resolution of 10~pm and a broad bandwidth of 200~nm~\cite{R26,R16}. However, this improved performance comes at the expense of a significantly larger footprint (7.03~mm$^2$) and reliance on thermo-optic tuning elements, which introduces additional operational complexity and increases time and power overhead.
\\\hspace*{1em}In summary, our proposed spectrometer can be implemented through one-shot image capture of the speckle pattern from a pure passive silicon photonic chip with a footprint of 2 mm$^2$. An experimental resolution of 10 pm and a bandwidth of 200 nm can be achieved, resulting in an ultrahigh bandwidth-to-resolution ratio of 20,000. Both narrowband and broadband spectra can be accurately reconstructed. This performance is primarily enabled by a large number of independent sampling channels, with approximately 2730 ($\rho_{thr}=0.5$) contributing to the spectral reconstruction. This enables finer spectral resolution and greater information throughput. While the current design occupies 2 mm$^2$, the footprint can be further reduced by densifying the waveguide layout or adopting more compact passive elements. The operational bandwidth of our spectrometer can be further enhanced by substituting the input grating coupler with a broadband I/O device. Since our design employs a pure passive photonic chip, the associated fabrication complexity and cost are low. In the future, the performance of the spectrometer can be further improved by targeting three key aspects: spectral resolution, bandwidth, and the number of independent sampling channels. Firstly, the spectral resolution can be enhanced by increasing the spatial decorrelation of the speckle patterns. This can be achieved by introducing more unbalanced MZIs with larger variability in optical path differences, thereby generating more diverse spectral interference patterns across image pixels. Secondly, the operational bandwidth can be significantly extended by addressing coupling limitations. In particular, replacing broadband grating couplers with polarization-insensitive I/O devices (e.g., edge couplers) reduces loss and ensures uniform coupling across wavelengths. Optimizing antenna design or fabricating the optical network on a silicon nitride platform can also enhance operational bandwidth. Finally, the number of independent sampling channels can be increased by maximizing speckle randomness and minimizing spatial redundancy. This can be achieved by enhancing the randomness of the interference network to produce highly decorrelated spatial outputs, enabling more efficient utilization of sensor pixels and improving the overall information capacity of the spectrometer. Collectively, these strategies pave the way for future implementations featuring higher resolution, broader operational bandwidth, and an increased number of independent sampling channels. 
\medskip
\\\noindent \textbf{Supporting Information} \par 
\noindent Supporting Information is available from the author.

\medskip
\noindent \textbf{Acknowledgments} \par 
\noindent This work was funded by the National Natural Science Foundation of China (No. 62305277), Natural Science Foundation of Guangdong Province (No.2024A1515012438), Guangzhou Association for Science and Technology (No. QT2024-011), and Guangzhou Science and Technology Bureau (No. 2024A04J4234). The authors acknowledge Applied Nanotools Inc. for photonic chip fabrication.

\medskip
\noindent \textbf{Conflict of interest} The authors declare no conflict of interest.

\medskip
\noindent \textbf{Data Availability Statement} The data that support the findings of this study are available from the corresponding author upon reasonable request.

\bibliographystyle{naturemag}
\bibliography{Arxiv/0_main_arxiv}

\renewcommand{\bibpreamble}{$^\ddag${Corresponding author: \textcolor{magenta}{yeyutong@hkust-gz.edu.cn}}}

\end{document}


\title[Article Title]{Supplementary Information for: Single-Shot Integrated Speckle Spectrometer with Ultrahigh Bandwidth-to-Resolution}


\author{Wenzhang Tian$^\ddagger$}
\author{Hao Chen$^\ddagger$}
\author{Mingyuan Zhang}
\author{Zengqi Chen}
\author{Yeyu Tong}
\email{yeyutong@hkust-gz.edu.cn}
\affiliation{Microelectronic Thrust, The Hong Kong University of Science and Technology (Guangzhou), 511453, Guangzhou, Guangdong, PR China}

\maketitle
\tableofcontents
\listoffigures
\newpage

\section{SIMULATION OF BROADBAND GRATING COUPLER AND ANTENNA}\label{ssec1}

\noindent{The} broadband grating coupler features a subwavelength grating structure combined with an adiabatic taper to achieve high coupling efficiency and ultra-broad bandwidth. As depicted in Figure S1a, the adiabatic taper smoothly transitions from a narrow waveguide width of 0.5 µm to a broader grating region width of 15 µm over a length of 400 µm. The grating region, shown in detail by the scanning electron microscope (SEM) image in Figure S1b, consists of periodic high and low refractive index regions with a subwavelength periodicity of 1.14 \textmu m in the x-direction and 0.425 \textmu m in the y-direction\cite{grating}. This configuration creates a uniform effective refractive index, which reduces waveguide dispersion and enhances the bandwidth. Each grating period contains segments with widths of 85 nm and 285 nm, corresponding to the high and low refractive index regions, with filling factors of 0.33 and 0.67, respectively. These filling factors are carefully optimized to balance bandwidth and coupling efficiency, ensuring robust performance across a broad spectral range. Additionally, the grating operates at an oblique incident angle of 15°, an angle specifically selected to optimize phase matching between the free-space mode and the guided mode. The grating, fabricated through a fully etched, single-step electron beam lithography (EBL) process, demonstrates high fabrication precision and uniformity, as shown in Figure S1b. Figure S1c and S1d, showing the near-field optical intensity at 1550 nm and the transmission spectrum respectively, demonstrate the broadband and efficient coupling performance of the input grating. Notably, these results were obtained through comprehensive 3D FDTD simulations, which confirm that a well-confined beam emerges from the grating region and reveal a relatively flat transmission (around –9 dB) from 1400 to 1550 nm, with a drop-off beyond 1575 nm.
\\The design of the antenna is illustrated in Figures S2a and S2b, where eight partially etched grating teeth are arranged in a compact, curved geometry with lateral and longitudinal extents of 5 µm and 10 µm, respectively. Each grating tooth is 220 nm thick, spaced at 548 nm intervals, and has a width of 182 nm, forming a periodic structure optimized for efficient directional radiation into free space. This compact footprint enables dense integration of multiple antennas on a single chip, as shown in the optical micrograph in Figure S2c. The near-field optical intensity distribution, calculated at a plane 1 µm above the antenna surface, is shown in Figure S2d. A distinct central lobe is observed, indicating strong and directional emission from the antenna structure. The corresponding far-field radiation profile, presented in Figure S2e, exhibits a main lobe centered along the surface normal, with a radiation angle concentrated within approximately ±12°, indicating a strong overlap with the collection cone of a typical 10$\times$ objective lens. This radiation pattern is specifically engineered to match the numerical aperture (NA) of the objective, and in our case, the emitted field lies well within the 0.22 NA of the 10$\times$ objective, facilitating efficient free-space coupling. Furthermore, the simulated transmission spectrum obtained from 3D FDTD analysis, shown in Figure S2f, exhibits a consistently high and flat response across the 1400–1600 nm wavelength range, with transmission above –6 dB over the entire band. This broadband performance confirms that the antenna design supports low-loss radiation and uniform optical coupling efficiency over a wide spectral range.

\section{MEASUREMENT METHOD}\label{ssec2}
The experimental setup is illustrated in Figure~S3. A tunable laser source (TLS, Santec TSL-570 and TSL-770) is used to sequentially scan discrete wavelengths across the 1400–1600~nm range during the pre-calibration phase. Specifically, the TLS steps through \(L = 2001\) wavelength points with a step interval of 0.1~nm and a time interval of 0.1~s per step. At each wavelength, light is coupled into the photonic integrated circuit (PIC) via a single-mode fiber (SMF), and the resulting speckle pattern \(I \in \mathbb{R}^{MN}\) is recorded by an infrared imaging system, which includes a 10× objective (MY10X-823), imaging lens (MVL12X3Z), and infrared camera (ARTCAM-991SWIR). The camera is configured with an exposure time of 0.5~ms, a frame rate of 100~fps, and a resolution of 512~×~512 pixels. Meanwhile, a commercial optical spectrum analyzer (OSA, Yokogawa AQ6780C) is used to simultaneously capture the reference spectrum. The total image acquisition time for the complete spectral range is approximately 200~s (\(0.1~\text{s} \times 2001\) steps), ensuring spectral sampling and accurate calibration of the wavelength-dependent transmission matrix.

Following calibration, unknown spectra are introduced for testing. Narrowband inputs are generated by the tunable laser, and broadband spectra are produced from an amplified spontaneous emission (ASE) source shaped using a programmable optical filter (WaveShaper 1000B). These signals are directed to the PIC through an optical switch (OS). The same infrared imaging system and configuration as used in the calibration stage is employed to capture the output speckle patterns in a single shot. For validation, the spectral response is simultaneously recorded using a commercial optical spectrum analyzer (OSA). The total acquisition and reconstruction time for a single unknown spectrum is approximately 1~s.

This experimental configuration facilitates the characterization and reconstruction of both narrowband and broadband spectra. The programmable optical filter allows for precise spectrum shaping, while the tunable laser provides fine control over the narrowband spectrum generation. The computer manages the data acquisition and spectral reconstruction, ensuring accurate and efficient spectral analysis.

\section{COMPUTATIONAL RECONSTRUCTION METHOD}
\noindent The reconstruction pipeline of our speckle-based spectrometer involves two main stages: calibration of the system transmission matrix $T$ and spectral reconstruction from a captured speckle image $\hat{S}$. Each stage incorporates algorithmic and theoretical components to ensure accurate and robust reconstruction under experimental noise.
\\\noindent\textbf{(1) Transmission Matrix Calibration:}  
We first calibrate the transmission matrix \( T \in \mathbb{R}^{MN \times L} \), where \( M\times N \) is the total number of pixels in the captured image and \(L\) is the number of discrete spectral sampling points. A tunable laser source (TLS) is swept across all wavelengths \( \{ \lambda_1, \lambda_2, \dots, \lambda_L \} \), and the corresponding speckle images are recorded. At each step, a one-hot encoded spectrum is injected, transformed to the DCT domain, and the matrix \( T \) is optimized by minimizing the least-squares error.

\begin{algorithm}[H]
\caption{Transmission Matrix Calibration}
\begin{algorithmic}[1]
\REQUIRE Monochromatic input spectrum $x$, tunable laser source (TLS), initial transmission matrix $T$
\ENSURE Calibrated transmission matrix $T$
\STATE \textbf{Step 1:} Initialize system matrix: $T \gets 0.5 \times \text{ones}(MN, L)$
\STATE \textbf{Step 2:} Loop over calibration wavelengths
\FOR{each wavelength $\lambda_i \in \{ \lambda_1, \lambda_2, \ldots, \lambda_L \}$}
    \STATE \hspace{1em} \texttt{setTLSWavelength}$(\lambda_i)$
    \STATE \hspace{1em} $x \gets \mathrm{DCT}(\texttt{oneHot}(i))$
    \STATE \hspace{1em} $I \gets \texttt{captureImage}()$
    \STATE \hspace{1em} Compute loss: $Loss \gets \| I - T x \|^2$
    \STATE \hspace{1em} Update transmission matrix: $T \gets T - \tau \nabla_T Loss$
\ENDFOR
\STATE \textbf{Step 3:} Return calibrated matrix
\RETURN $T$
\end{algorithmic}
\end{algorithm}

\noindent\textbf{(2) Spectrum Reconstruction:} 
Once the transmission matrix \( T \in \mathbb{R}^{MN \times L} \) is calibrated, we reconstruct the unknown input spectrum from a single-shot speckle image \( I \in \mathbb{R}^{MN} \) captured by the infrared camera. The reconstruction process proceeds in three steps. Firstly, an initial estimate of the spectrum ${\hat{S}}$ is obtained using the pseudo-inverse of the transmission matrix, i.e., \( {\hat{S}} \gets T^{\dagger} I \). This provides a fast but coarse reconstruction. Secondly, to improve accuracy and suppress noise, a nonlinear optimization is applied by minimizing the least-squares loss \(\min\| I - T {\hat{S}} \|^2\) through iterative gradient descent. Finally, if spectral sparsity is assumed, the solution is projected into the discrete cosine transform (DCT) domain for compressed sensing recovery. The DCT transformation improves noise resilience by concentrating signal energy into a small number of coefficients, facilitating sparse reconstruction and denoising. The final spectrum is obtained by applying the inverse DCT to the optimized coefficients. This three-step procedure enables robust and accurate spectrum recovery from high-dimensional speckle data.

\begin{algorithm}[H]
\caption{Spectrum Reconstruction}
\begin{algorithmic}[1]
\REQUIRE Calibrated transmission matrix $T \in \mathbb{R}^{MN \times L}$, captured speckle image $I \in \mathbb{R}^{MN}$
\ENSURE Reconstructed spectrum ${\hat{S}} \in \mathbb{R}^{L}$
\STATE \textbf{Step 1:} Initial estimate via pseudo-inverse: ${\hat{S}} \gets T^\dagger I$
\STATE \textbf{Step 2:} Nonlinear least-squares optimization:
\FOR{$j = 1$ to $T_{\text{max}}$}
    \STATE \hspace{1em} Compute loss: $Loss \gets \| I - T {\hat{S}} \|^2$
    \STATE \hspace{1em} Compute gradient: $\nabla_{\hat{S}} L \gets -2 T^\top (I - T {\hat{S}})$
    \STATE \hspace{1em} Gradient descent update: ${\hat{S}} \gets {\hat{S}} - \tau \nabla_{\hat{S}} L$
\ENDFOR
\STATE \textbf{Step 3:} DCT-domain recovery:
\STATE \hspace{1em} $\hat{S}_{\text{DCT}} \gets {\hat{S}}$, then $\hat{S} \gets \mathrm{IDCT}(\hat{S}_{\text{DCT}})$
\RETURN $\hat{S}$
\end{algorithmic}
\end{algorithm}

\noindent\textbf{Theoretical Motivation for DCT-based Recovery:}  
The following provides theoretical motivation and key advantages of incorporating DCT domain processing into the reconstruction pipeline, as described in Algorithm~2. While the algorithm outlines the implementation steps, this section summarizes the mathematical rationale behind its design\cite{DCT}. Without DCT, direct inversion \(\hat{S} = T^{-1} I\) is highly sensitive to experimental noise. Nonlinear optimization mitigates this by minimizing the least-squares loss:
\begin{equation}
\varepsilon = \left\| I - T \hat{S} \right\|^2
\end{equation}
However, for complex or broadband spectra, where the signal is not sparse in the original domain, this approach is prone to noise amplification and limited denoising ability.To address this, we introduce the discrete cosine transform (DCT), which converts the spectrum into a sparse representation, thereby making compressed sensing (CS) techniques applicable. The transformed system minimizes:
\begin{equation}
\varepsilon_{\text{DCT}} = \left\| I - T_{\text{DCT}} \hat{S}_{\text{DCT}} \right\|^2
\end{equation}
where \(T_{\text{DCT}} = T D^{-1}\), \(\hat{S}_{\text{DCT}} = D S\), and \(D\) is the DCT basis matrix.
As shown in Figure~S4, spectral reconstructions without applying DCT-based processing suffer from significant noise and distortion for both narrowband and broadband inputs. Specifically, Figure~S4a and S4c show that the relative errors reach \(\varepsilon = 0.154\) and \(\varepsilon = 0.325\), respectively, indicating poor fidelity and high sensitivity to noise. In contrast, when DCT-domain optimization is employed, the reconstructed spectra in Figure~S4b and S4d exhibit excellent agreement with the reference, with relative errors reduced to \(\varepsilon = 0.041\) and \(\varepsilon = 0.035\), respectively. These results clearly demonstrate that DCT-domain processing not only enhances reconstruction robustness against noise but also significantly improves accuracy across diverse spectral profiles.

\noindent

\section{ESTIMATION OF SAMPLING CHANNELS}

\noindent To quantitatively assess the number of statistically independent spatial sampling channels in the random speckle images, we developed an approach that integrates both spatial and spectral information. Specifically, our experimental data consist of $L$ speckle images, each with spatial dimensions $M \times N$, acquired at distinct wavelengths. These $L$ images are concatenated into a composite matrix: $I \in \mathbb{R}^{MN \times L}$, where each column represents a flattened spatial image corresponding to a specific wavelength channel. The matrix $I$ thus encapsulates the complete spatial-spectral response of the speckle patterns across the entire measured spectrum.

To evaluate spatial independence among sampling points, we first select the spatially central pixel of the composite speckle image as a reference pixel, denoted by $(x_0, y_0)$. We then compute the Pearson correlation coefficient $C(x,y)$ between the spectral response at the reference pixel $I(x_0,y_0,\lambda)$ and the spectral response at an arbitrary neighboring pixel $I(x,y,\lambda)$ within a spatial radius $r$:
\begin{equation}
C(x,y) = \frac{\mathrm{cov}[I(x_0,y_0,\lambda), I(x,y,\lambda)]}{\sigma[I(x_0,y_0,\lambda)]\,\sigma[I(x,y,\lambda)]}
\end{equation}
where $\mathrm{cov}[\cdot,\cdot]$ denotes covariance, and $\sigma[\cdot]$ is the standard deviation across spectral channels. 
Next, to characterize the spatial decay of correlation, we group neighboring pixels into concentric annuli based on their distance from the reference pixel, and compute the average absolute value of \(C(x,y)\) within each ring. This averaging yields the radial correlation function, \(\rho(r)\), which is defined as:
\begin{equation}
\rho(r) \;=\; \text{Average}\bigl(C(x,y)\;\text{for}\;\Vert(x,y)-(x_0,y_0)\Vert \approx r\bigr)
\end{equation}
where $\rho(r)$ represents how correlation decays with increasing distance from the selected center, as shown in Figure S5a and S5b. Two pixels separated by a distance r are considered statistically independent when the radially averaged correlation $\rho(r)$ falls below a threshold $\rho_{thr}$.
We then define the minimum uncorrelated radius \(r_{\min}\) as the smallest radius \(r\) for which \(\rho(r)\) drops below a prescribed threshold \(\rho_{\mathrm{thr}}\):
\begin{equation}
r_{\min} \;=\; \min\{\,r \mid \rho(r)\;<\;\rho_{\mathrm{thr}}\}
\end{equation}
Finally, we estimate the total number of statistically independent sampling channels by assuming each independent unit occupies a square area of side length \(2r_{\min}\). Dividing the total image area \(\,(M\times N)\) by this unit area thus gives\cite{redding2013compact}:
\begin{equation}
\text{Sampling channels} \;=\; \frac{M \times N}{\bigl(2r_{\min}\bigr)^2}
\end{equation}
Figure~S5c illustrates how varying \(\rho_{\mathrm{thr}}\) from 0.3 to 0.7 alters \(r_{\min}\), causing the estimated number of sampling channels to rise from about 1364 to 6419. Specifically, we obtain approximately 2730 independent channels at \(\rho_{\mathrm{thr}} = 0.5\).

\clearpage
\begin{figure*}[htbp]
  \centering{
  \includegraphics[width = 1\linewidth]{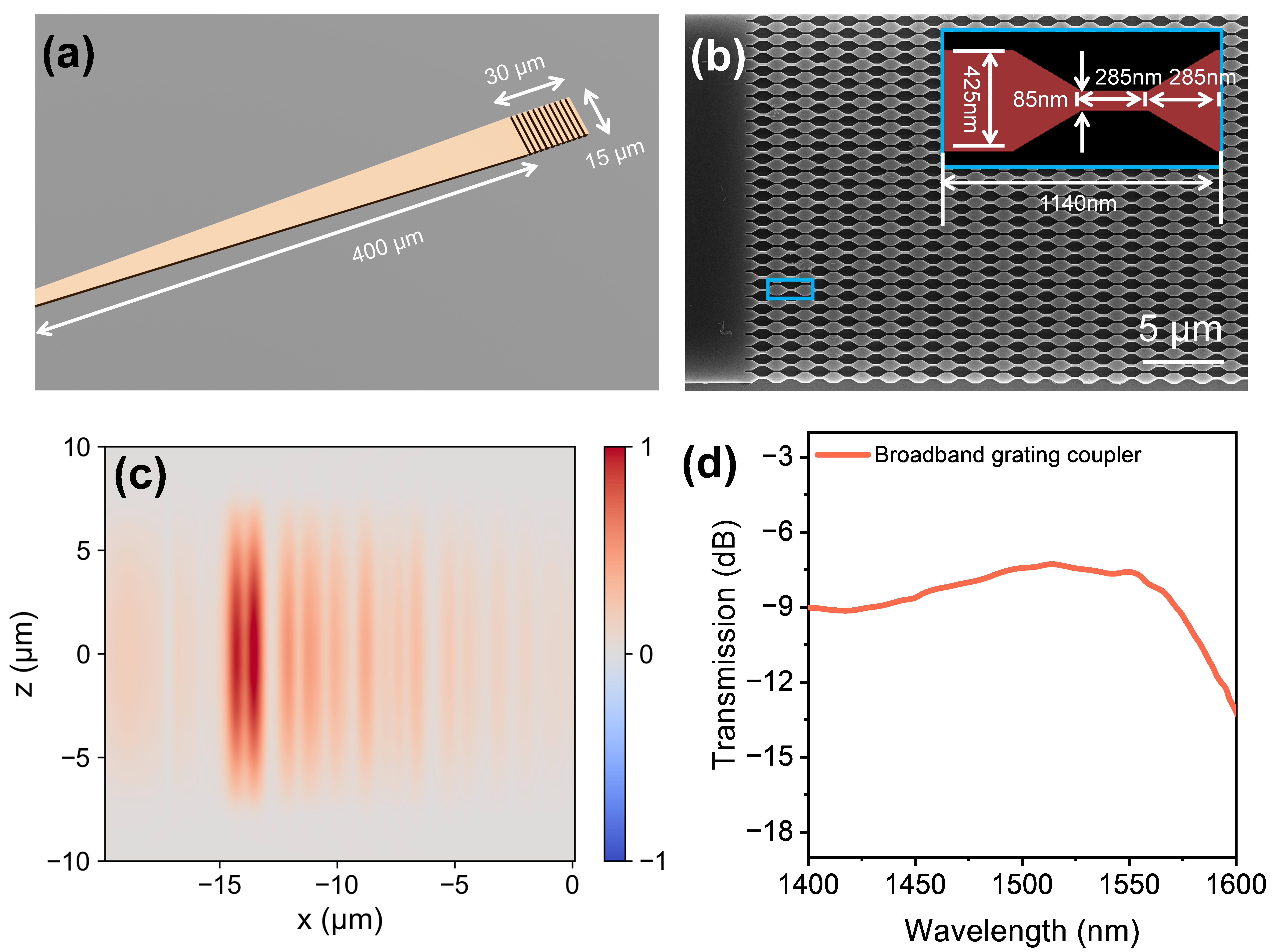}
  } 
  \caption{Schematic representation of the broadband grating optimized for 1550 nm. (a) 3-D model of the silicon-based broadband grating coupler. (b) Scanning electron microscope (SEM) image of the broadband grating coupler. (c) Calculated near-field optical intensity distribution. (d) Simulated transmission of the broadband grating coupler from 1400 nm to 1600 nm.}
 \label{FigureS1}
\end{figure*}

\clearpage
\begin{figure*}[htbp]
  \centering
  \includegraphics[width = 1\linewidth]{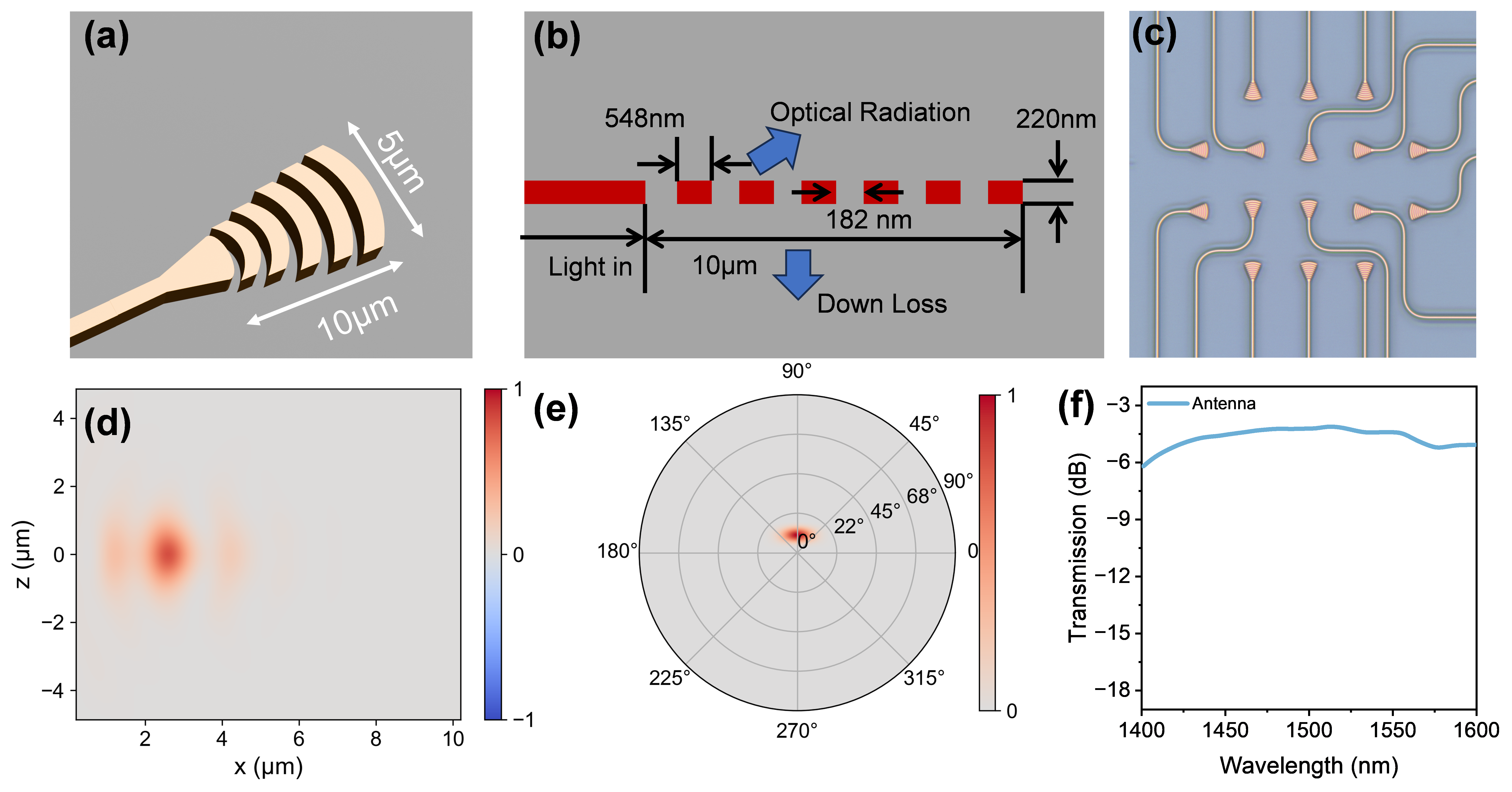}
  \caption{Schematic representation of the $10 \times 5$ $\mu m^2$ antenna optimized for 1550 nm. (a) 3-D model of the silicon-based antenna. (b) Design parameters obtained for the optimized antenna. (c) Microscope image of antenna array. (d) Calculated near-field optical intensity distribution. (e) Calculated far-field optical intensity distribution at 1550 nm wavelength. (f) Simulated transmission of nano antenna from 1400 nm to 1600 nm.}
 \label{FigureS2}
\end{figure*}

\clearpage
\begin{figure*}[htbp]
  \centering{
  \includegraphics[width = 1\linewidth]{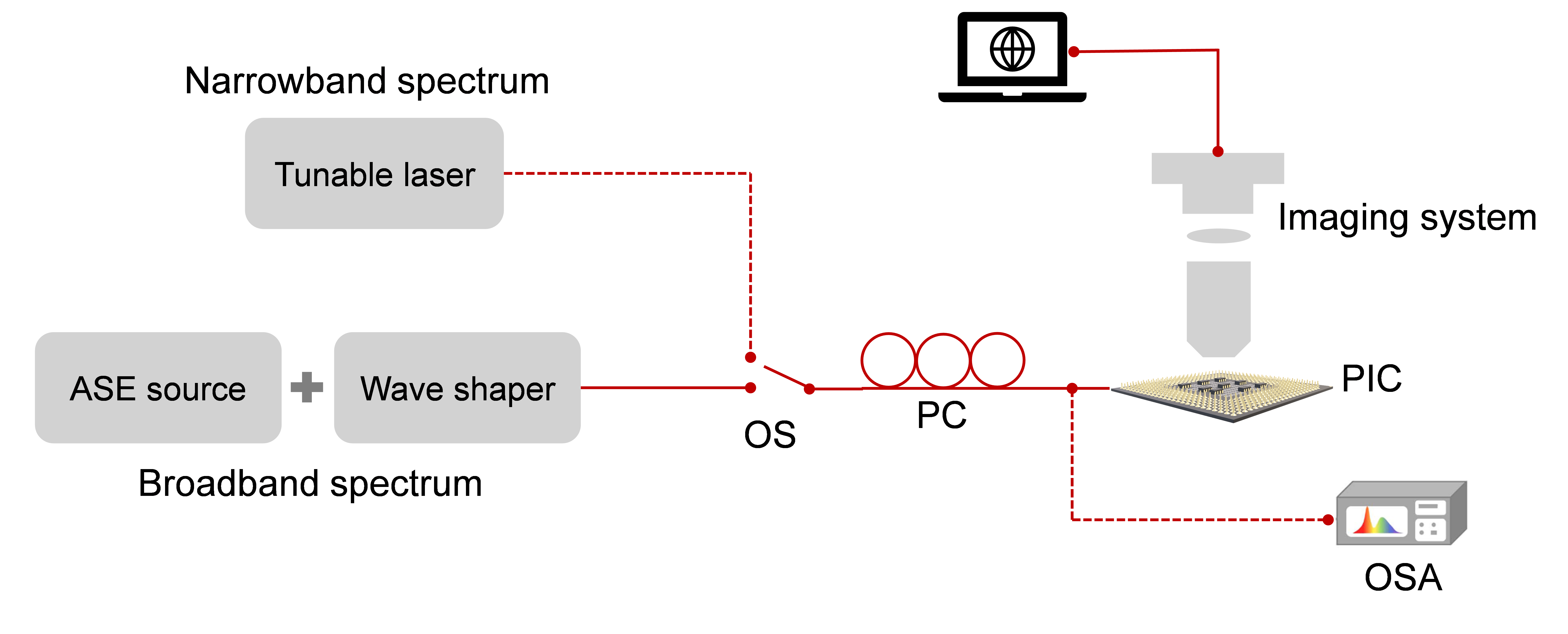}
  } 
  \caption{Diagram of the experimental setup. OS: Optical switch. PC: Polarization controller. PIC: Photonics integrated circuits. OSA: Optical spectrum analyzer. }
 \label{FigureS3}
\end{figure*}

\clearpage
\begin{figure*}[htbp]
  \centering
  \includegraphics[width = 1\linewidth]{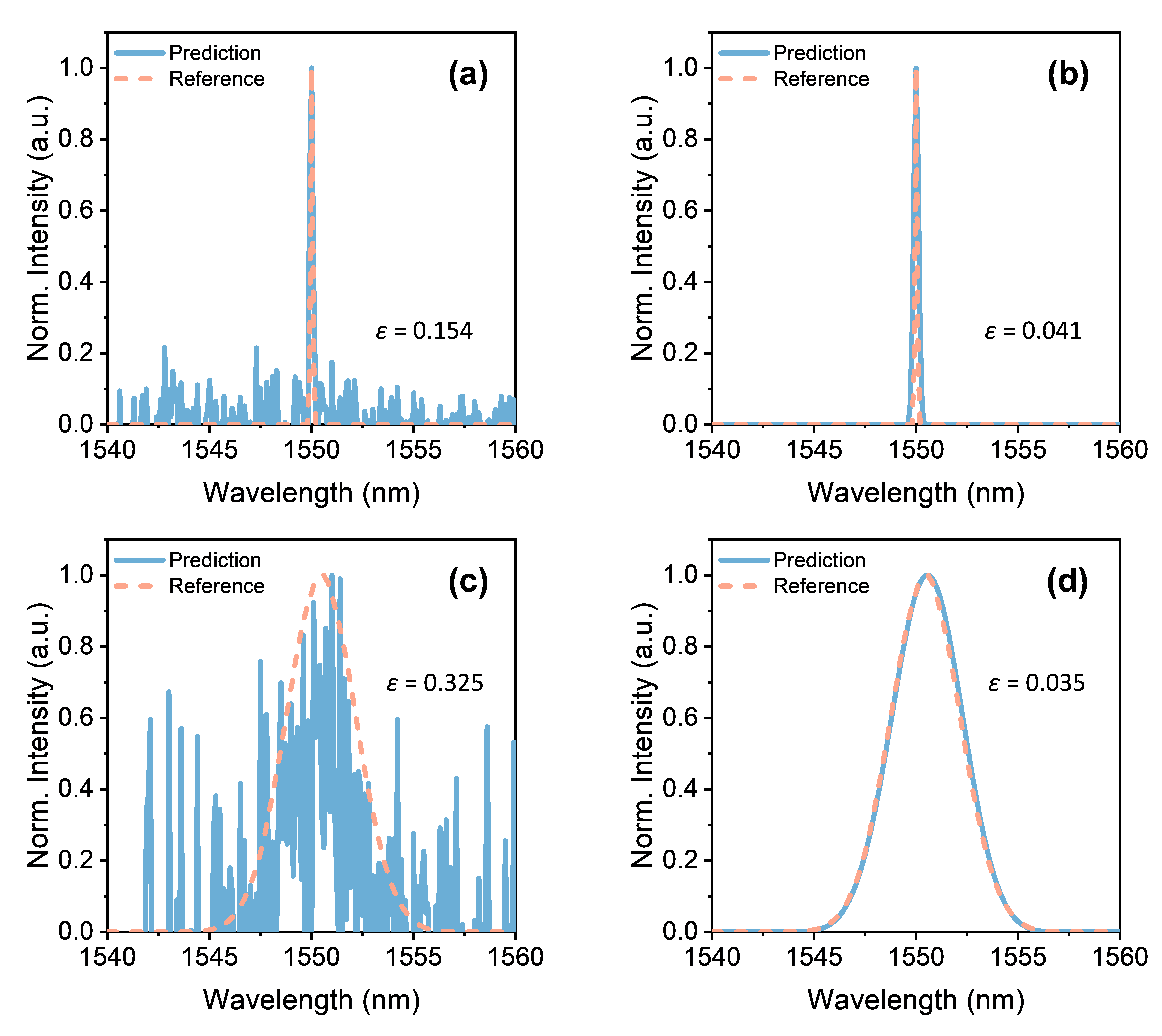}
  \caption{(a) Reconstructed narrow peak with a FWHM of 10 pm. (b) Reconstructed narrow peak with a FWHM of 10 pm by nonlinear optimization in the DCT domain. (c) Reconstructed continuous spectrum with a FWHM of 5 nm. (d) Reconstructed continuous spectrum with a FWHM of 5 nm by nonlinear optimization in the DCT domain.}
 \label{FigureS2}
\end{figure*} 

\clearpage
\begin{figure*}[htbp]
  \centering{
  \includegraphics[width = 1.0\linewidth]{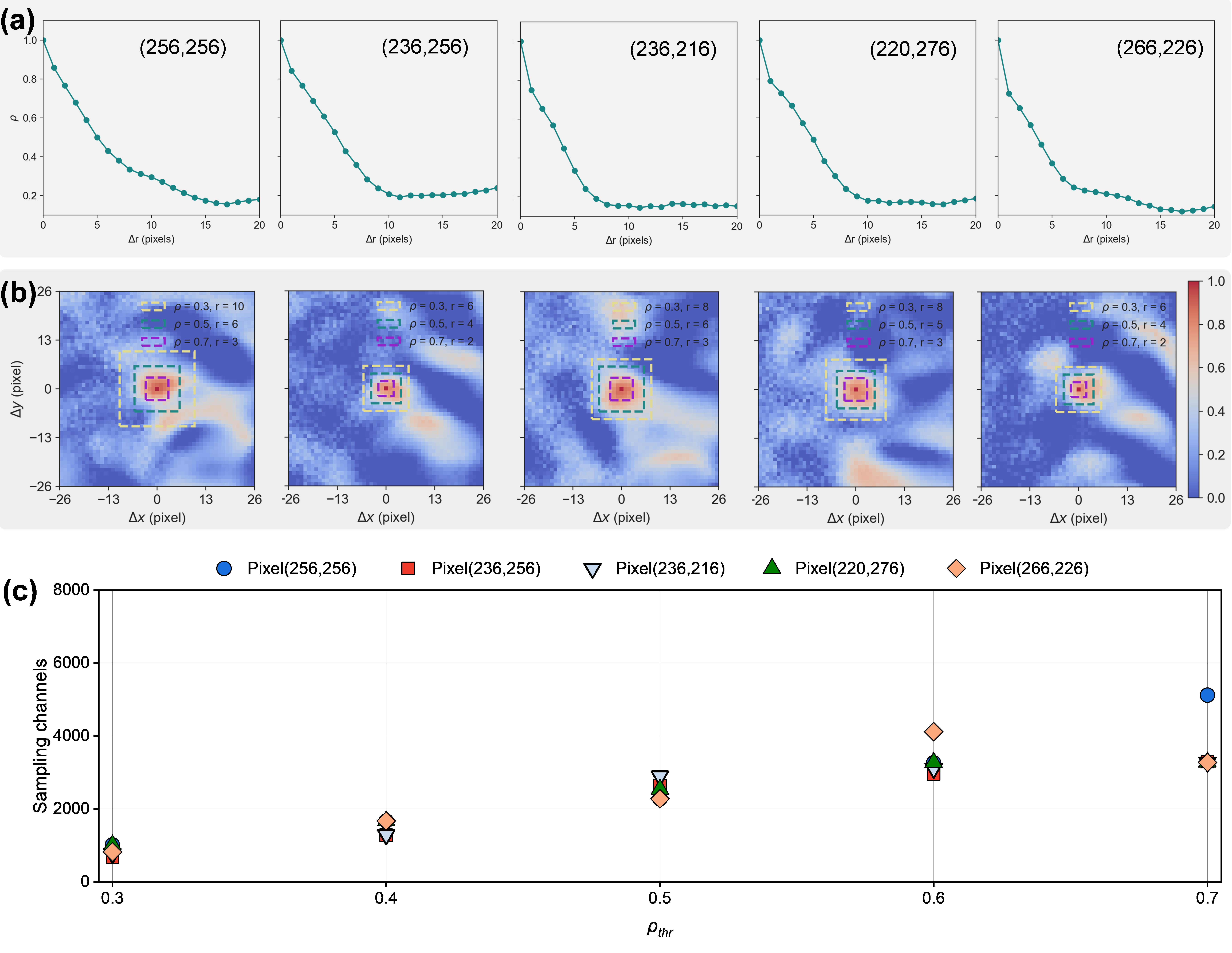}
  } 
  \caption{Estimation of spatial correlation and independent sampling channels. (a) Radial correlation profiles $\rho(r)$ computed for different selected central pixels, used to estimate the minimum uncorrelated radius $r_{\mathrm{min}}$. (b) Pearson correlation maps between each selected central pixel and its surrounding pixels. Colored boxes indicate the estimated minimum independent sampling units under thresholds $\rho_{\mathrm{thr}} = 0.3,\ 0.5,\ 0.7$, with side lengths equal to $2r_{\mathrm{min}}$. (c) Estimated number of statistically independent sampling channels across the full spectral range (1540–1560\,nm) as a function of $\rho_{\mathrm{thr}}$, for different central pixel selections. Specifically, the average number of independent sampling channels is approximately 2730 at $\rho_{\mathrm{thr}} = 0.5$.}
 \label{FigureS4}
\end{figure*}

\clearpage
\bibliographystyle{naturemag}
\bibliography{Arxiv/0_SI_arxiv}




